\documentclass[epjCONF,columns]{svjour} 
\usepackage{graphics}
\usepackage[varg]{txfonts} 
\usepackage[latin1]{inputenc}
\session-title{Hadron Collider Physics Symposium 2011}
\begin{document}
\title{$t\bar{t}$ pair production cross section measurement at the LHC}
\author{Tae Jeong Kim\thanks{\email{Tae.Jeong.Kim@cern.ch}} for the ATLAS and CMS collaborations}
\institute{Korea University, Seoul}
\abstract{
Measurement of $t\bar{t}$ pair production cross sections 
with an integrated luminosity of around \mbox{1 fb$^{-1}$}
at $\sqrt{s}$ = 7 TeV obtained with the ATLAS and CMS detectors are reported.
The inclusive cross sections in dilepton (ee, $e\mu$, $\mu\mu$ and $\mu\tau$), lepton+jets (e, $\mu$) 
and all hadronic decay modes are measured. In addition to inclusive cross section measurement, 
the study of jet multiplicity with additional jets are also presented, which is important to constrain the initial state radiation. 
Measurement of the charge asymmetry at the LHC is also presented. 
All measurements are compatible with Standard Model predictions. 
} 
\maketitle
\section{Introduction}
\label{intro}
The Large Hadron Collider (LHC) accumulated the data corresponding to an integrated luminosity
of almost 1 fb$^{-1}$ in both experiments ATLAS~\cite{ATLASdet} and CMS~\cite{CMSdet} by the summer in 2011 for HCP2011. 
At the LHC, the $t\bar{t}$ production cross section at $\sqrt{s}$ = 7 TeV is
predicted to be 164.6 pb by approximate next-to-next-leading-order (NNLO) calculation and 157.5 pb 
by next-leading-order (NLO) calculation.
The cross section measurement of $t\bar{t}$ is important for 
testing the perturbative QCD which is successful so far 
and searching for new physics. Any deviation would indicate possible new physics.
It is crucial to measure the cross section in all decay modes 
since new physics can appear in any different decay modes.
Top quark decays almost exclusively to W boson ans $b$ quark. 
Therefore, the decay mode entirely depending on W boson branching ratio.
In dilepton decay mode (ee, $e\mu$, $\mu\mu$), tau leptonic decay is included.
Considering $Br(\tau \to l\nu_l\nu_\tau)$ = 0.35, 
the branching ratio would be 6.8\%, 3.8\%, 30\% and 44\% for dilepton, lepton+tau, lepton+jets and all hadronic decay modes, respectively.
In addition to inclusive cross section,       
the jet multiplicity distribution of $t\bar{t}$ with additional jets in lepton+jets decay mode is shown, which is 
important measurement to constrain the initial state radiation (ISR). 
As the deviation has been observed by Tevatron, the charge asymmetry measurement at the LHC
are also performed using the fact that the width of top quark is slightly broader than anti-top quark in
rapidity distribution at the LHC. 

\section{Samples \& Objects}
\label{sec:sample}
At ATLAS, MC@NLO is interfaced with HERWIG (Parton Showering) and JIMMY (Underlying Events).
Approximate NNLO of 164.6 pb is used for normalization.
At CMS, the signal sample of $t\bar{t}$ is modeled by MADGRAPH with PYTHIA matching up to three
additional partons. NLO cross section of 157.5 pb is used for normalization.
In both experiments, the $\tau$ decay is handled by TAUOLA and 
top quark mass is assumed to be \mbox{172.5 GeV/c$^{2}$}.

In top quark analysis, almost all physics objects are used.
At ATLAS, the absolute pseudo-rapidity ($|\eta|$) of electrons and muons are 
required to be within 2.5.
Taus are reconstructed using Boosted Decision Tree.
Calo-jets are reconstructed using anti-kt algorithm with R=0.4.
Missing transverse energy ($E^{miss}_{T}$) is calculated using 
the opposite direction of vector sum of jets, electrons, muons and unclustered calorimeter energy.
At CMS, physics objects are reconstructed through particle-flow reconstruction algorithm
which combines all information from all sub-detectors and
reconstruct all particles.
The $|\eta|$ of electrons and muons are
required to be within 2.5 and 2.4, respectively. 
Taus are reconstructed using Hadron plus strips algorithm.
Particle-flow jets are reconstructed using anti-kt algorithm with R=0.5.
$E^{miss}_{T}$ is the opposite direction of vector sum of reconstructed particles.
\section{Cross section measurements}
\label{sec:inclus}
\subsection{Dilepton ({\rm ee, e$\mu$, $\mu\mu$})}
\label{sec:dilepton}
\begin{figure}
\begin{center}
\resizebox{0.95\columnwidth}{!}{%
\includegraphics{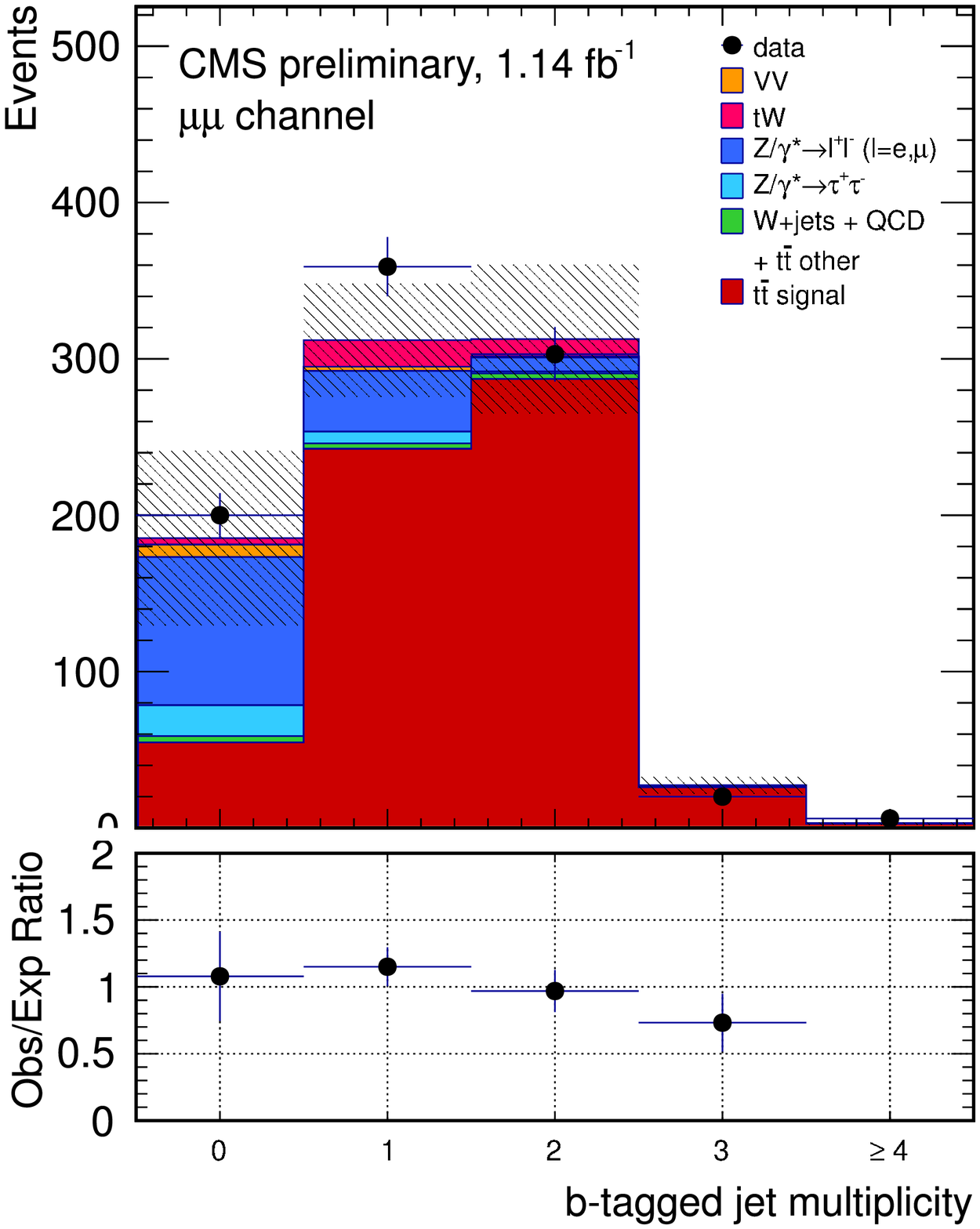}
\includegraphics{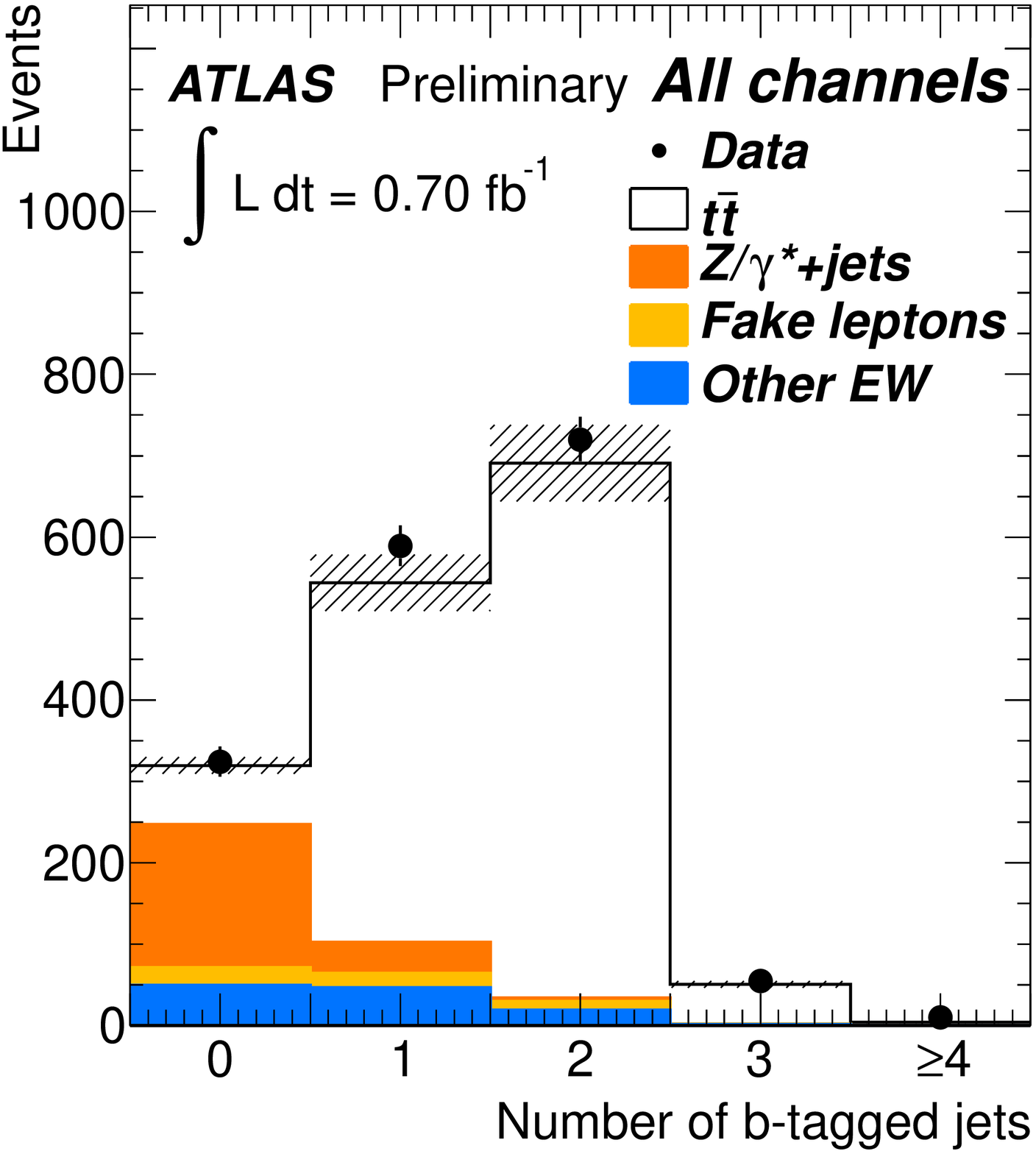}
}
\caption{Distributions of b-tagged jet multiplicity in $\mu\mu$ at CMS (left) and in all decay modes at ATLAS (right) 
after applying final selections except b-tagging.}
\label{fig:dilepton}       
\end{center}
\end{figure}

The dilepton decay mode (ee, e$\mu$, $\mu\mu$) provides clean signals by requiring two isolated leptons with two jets and 
$E^{miss}_{T}$ even though the branching ratio is small.
ATLAS performed the analysis with the integrated luminosity of \mbox{0.7 fb$^{-1}$} with and without b-tagging separately~\cite{CERNATLASdilepton}.
The invariant mass of dilepton must be above \mbox{15 GeV} to 
remove multi-jet event sample which does not describe well low mass region.
Z boson veto requiring \mbox{$|M_{{\it ll}}-M_{Z}|$ $>$ 10 GeV} and \mbox{$E^{miss}_{T}$ $>$ 60 GeV} (or \mbox{40 GeV} with b-tagging) 
to remove multi-jet events are applied for $ee$ and $\mu\mu$ decay modes. 
Additionally \mbox{$H_T$ $>$ 130 GeV} (or \mbox{140 GeV} with b-tagging) is applied.
Lepton efficiencies are obtained with $Z$ boson candidates in a data-driven way.
The Drell-Yan and multi-jet backgrounds are estimated using $Z$ mass window and Matrix method, respectively.
The cross section is obtained from the profile likelihood fitting.
The measured cross section without b-tagging is found to be
\begin{center}
$\sigma_{t\bar{t}}$=177$\pm$6(stat.)$\pm^{17}_{14}$(syst.)$\pm$8(lumi.) pb 
\end{center}
and with b-tagging
\begin{center}
$\sigma_{t\bar{t}}$=183$\pm$6(stat.)$\pm^{18}_{14}$(syst.)$\pm^{8}_{7}$(lumi.) pb.
\end{center}
CMS performed the analysis with the integrated luminosity of \mbox{1.1 fb$^{-1}$}~\cite{CERNCMSdilepton}.
The invariant mass of dilepton must be above \mbox{12 GeV} and 
\mbox{$|M_{{\it ll}}-M_{Z}|$ $>$ 15 GeV}. \mbox{$E^{miss}_{T}$ $>$ 30 GeV} is required for $ee$ and $\mu\mu$ decay modes. 
At least one b-tagging is applied.
Lepton efficiencies, Drell-Yan and multi-jet backgrounds are obtained in a data-driven way similar to ATLAS.
Distributions of b-tagged jet multiplicity at CMS and ATLAS are shown in Fig.~\ref{fig:dilepton}.
The cross section for each decay mode is obtained using counting method.
The combined result of the three decay modes is found to be
\begin{center}
$\sigma_{t\bar{t}}$=169.9$\pm$3.9(stat.)$\pm$16.3(syst.)$\pm$7.6(lumi.) pb
\end{center}
using the Best Linear Unbiased Estimator (BLUE) method.
The measured cross sections are consistency with NNLO prediction.
The systematic uncertainty is now dominant compared to the
result based on previous data at 
ATLAS~\cite{RefATLASdilepton3}-\cite{RefATLASdilepton36} and CMS~\cite{RefTop-10-001}-\cite{RefTop-11-002}.
\subsection{Dilepton ({\rm $\mu\tau$})}
\label{sec:ltau}
\begin{figure}
\begin{center}
\resizebox{0.9\columnwidth}{!}{%
\includegraphics{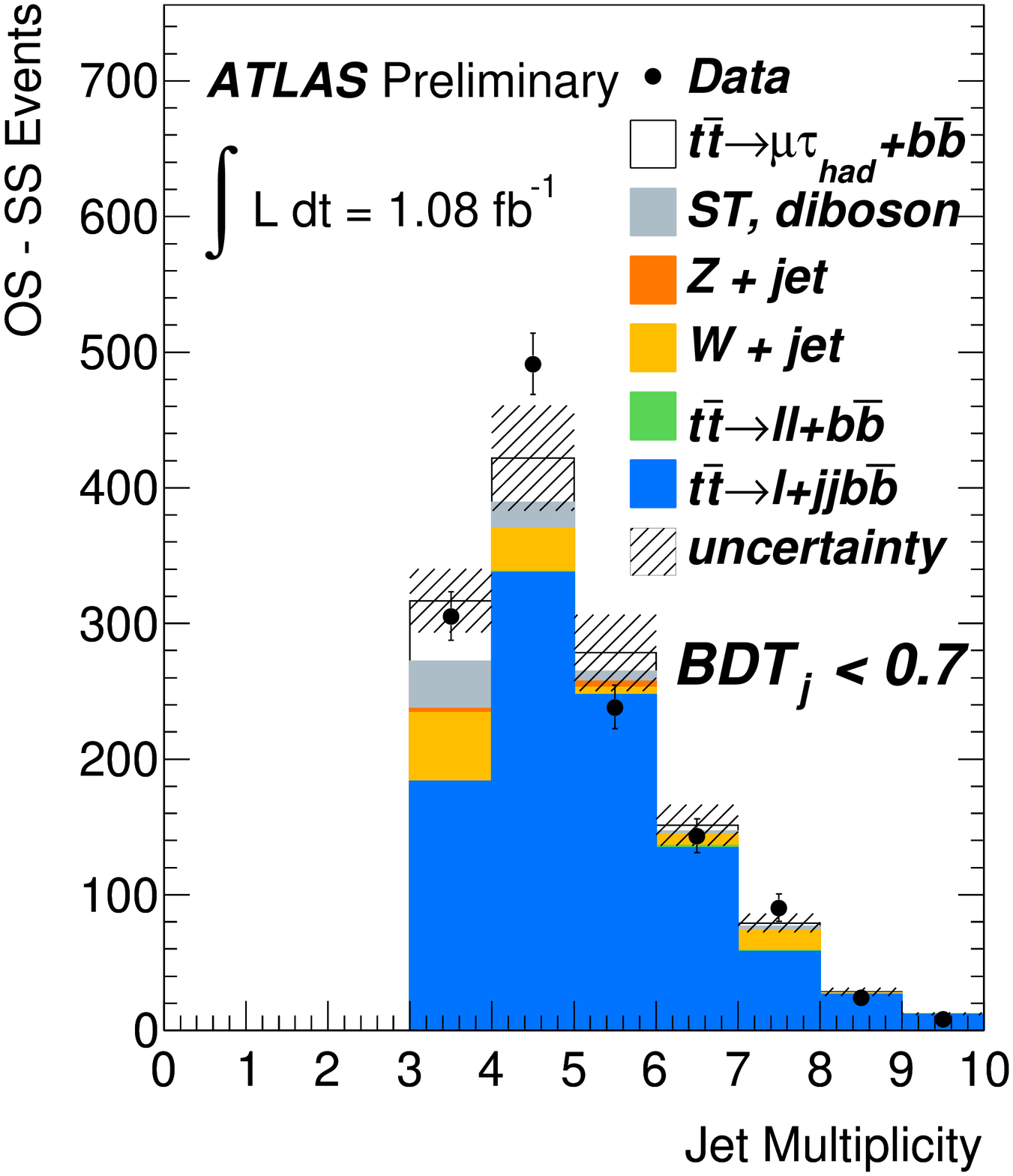}
\includegraphics{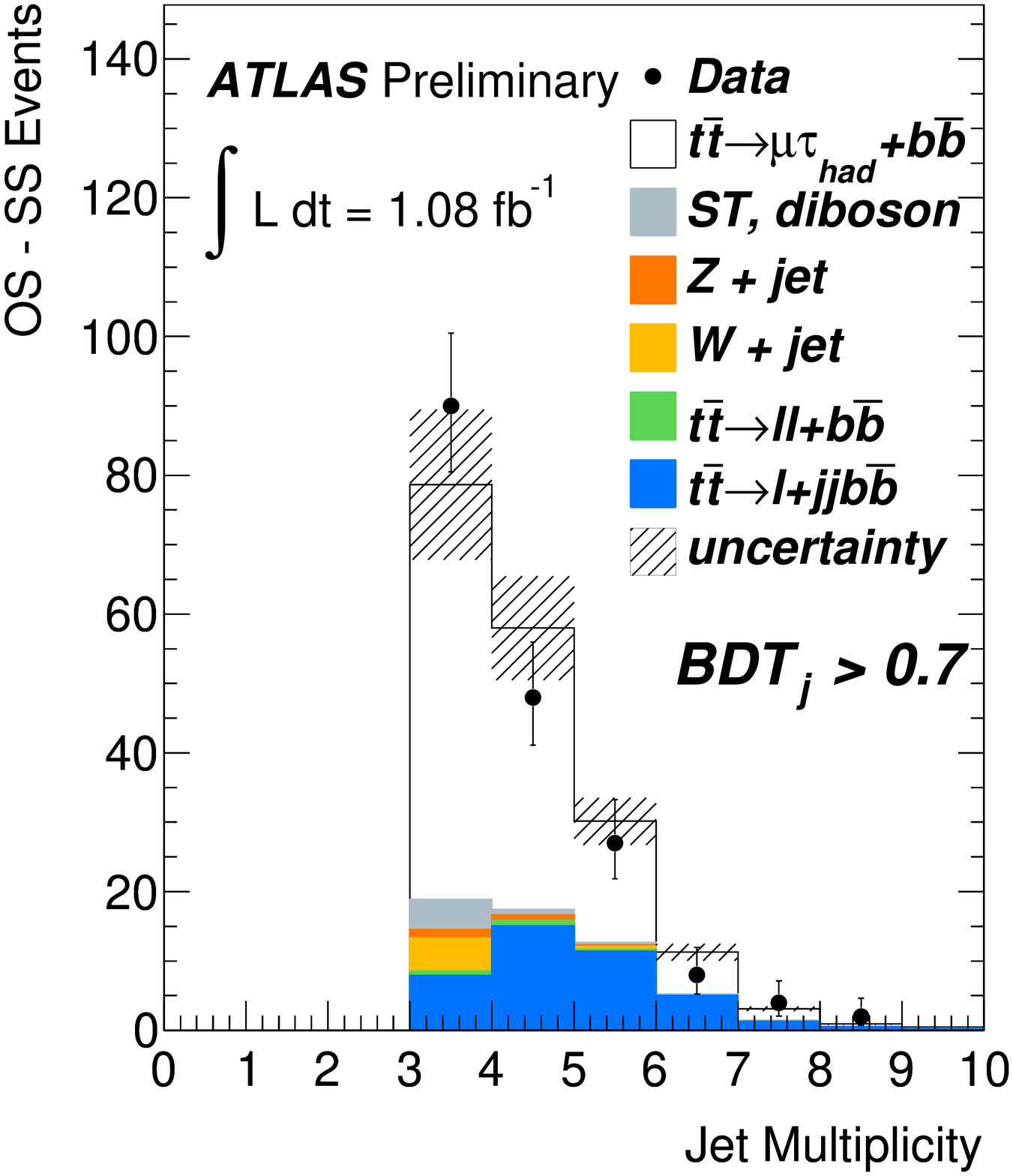}
}
\caption{The distribution of jet multiplicity subtracting same sign events after b-tagging at ATLAS.
(a) $BDT < 0.7$, (b) $BDT > 0.7$}
\label{fig:ATLAStau}       
\end{center}
\end{figure}

The $\mu+\tau$ decay mode is interesting since
the charged higgs can decay with the same topology when the higgs mass is larger than
top mass. Any deviation on cross section would indicate the new physics. Therefore,
reducing the systematic uncertainty is crucial in this decay mode.  
ATLAS performed the analysis with an integrated luminosity of \mbox{1.08 fb$^{-1}$}~\cite{CERNATLASmutau}.
This analysis follows lepton+jet analysis event selection since hadronic tau decay is considered.
Two tau candidates $\tau$ with 1 track and $\tau$ with more than 1 track are identified
using Boosted Decision Tree (BDT). 
The multi-jet events are removed by subtracting same sign events.
Distributions of jet multiplicity with \mbox{BDT $<$ 0.7} and \mbox{BDT $>$ 0.7} 
are shown in Fig.~\ref{fig:ATLAStau}.
Measured cross section at ATLAS is
\begin{center}
$\sigma_{t\bar{t}}$=142$\pm$21(stat.)$\pm^{20}_{16}$(syst.)$\pm$5(lumi.) pb.
\end{center}
CMS performed the analysis with an integrated luminosity of \mbox{1.1 fb$^{-1}$} data~\cite{CERNCMSmutau}.
Tau is identified with Hadrons plus strips (HPS) algorithm combining charged hadrons and 
calorimeter information in strips to take into account $\pi^{0}$. 
The fake rate from jets is estimated from multi-jet (gluon jet) and W+jets (quark jet) data sample.
Measured cross section is
\begin{center}
$\sigma_{t\bar{t}}$=148.7$\pm$23.6(stat.)$\pm$26.0(syst.)$\pm$8.9(lumi.) pb.
\end{center}
Main systematic uncertainties are from 
tau fake background estimation, identification and b-tagging efficiency.
\subsection{Lepton+jets}
\label{sec:ljets}
\begin{figure}
\begin{center}
\resizebox{0.8\columnwidth}{!}{%
\includegraphics{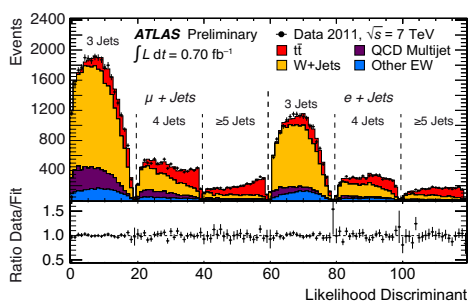} }
\caption{Result of combined fit to data in the exclusive three-jet bin, the exclusive four-jet bin and the inclusive five-jet bin of 
the e+jets and $\mu$+jets decay modes at ATLAS.}
\label{fig:ATLASljets}       
\end{center}
\end{figure}

In lepton+jet decay mode (e, $\mu$), the signature of a final state is one exclusive lepton, 4 jets and $E^{miss}_{T}$.
ATLAS performed the analysis with the integrated luminosity of \mbox{0.7 fb$^{-1}$}~\cite{CERNATLASleptonjets}.
Exclusively one isolated muon or electron must have \mbox{$p_T$ $>$ 20 } or 25 GeV, respectively. 
$E^{miss}_{T}$ is required to be larger than 35 and \mbox{25 GeV} for e and $\mu$, respectively. 
$M_T^W$ (Transverse mass of W boson) must be larger than \mbox{25 GeV} for electron channel and 
the sum of $M_T^W$ and $E^{miss}_{T}$ should be larger than \mbox{60 GeV} for muon channel to remove
further multi-jet contribution. 
The multi-jet shapes are obtained from data directly using Matrix method.    
The binned profile likelihood fitting is applied 
to likelihood discriminant which is as a function of lepton $\eta$, highest jet
$p_T$, event aplanarity and $H_T$. The result of the fit is shown in Fig.~\ref{fig:ATLASljets}.
Main systematic uncertainties are from signal MC generator, jet energy scale (JES), and ISR/FSR.
Measured cross section is found to be
\begin{center}
$\sigma_{t\bar{t}}$=179.0$\pm$3.9(stat.)$\pm$9.0(syst.)$\pm$6.6(lumi.) pb.
\end{center}
CMS performed the analysis with the integrated luminosity of \mbox{1.1 fb$^{-1}$} ($\mu$) and \mbox{0.8 fb$^{-1}$} (e)~\cite{CERNCMSleptonjets}.
Exclusively one isolated muon or electron must have \mbox{$p_T$ $>$ 35} or 45 GeV, respectively. 
$E^{miss}_{T}$ is required to be larger than 20 and \mbox{30 GeV} for e and $\mu$, respectively. 
b-tagging with secondary vertex algorithm is applied at CMS. 
The multi-jet shapes are obtained from data directly using non-isolated data.
Binned profile likelihood fitting is applied to
secondary vertex mass distribution in 1 b-tag and 2 b-tag jet bins.
Measured cross section is found to be
\begin{center}
$\sigma_{t\bar{t}}$=164.4$\pm$2.8(stat.)$\pm$11.9(syst.)$\pm$7.4(lumi.) pb.
\end{center}
Main systematic uncertainties are from $W$+jets Q$^2$ scale, b-tagging efficiency and JES.
Comparing the result with 36 pb$^{-1}$ in 2010~\cite{RefTop-10-002}-\cite{RefTop-10-003}, 
the statistical uncertainty is by far reduced and the systematic uncertainty is dominant. 
\subsection{Hadronic decay}
\label{sec:hdecay}
\begin{figure}
\begin{center}
\resizebox{0.75\columnwidth}{!}{%
\includegraphics{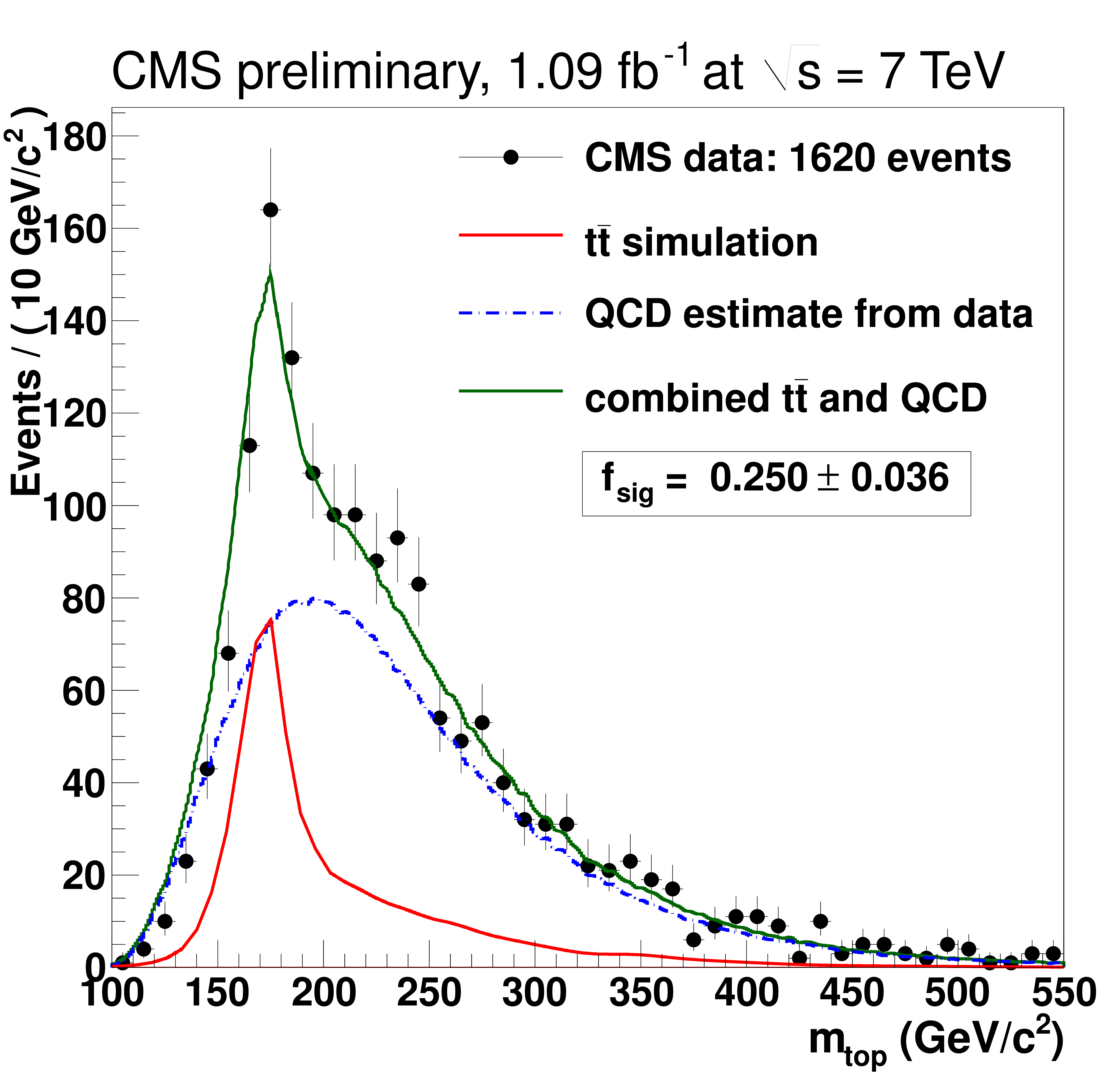}
}
\caption{Result of the fit to the reconstructed top mass distribution in hadronic decay mode at CMS.}
\label{fig:hadronic}       
\end{center}
\end{figure}

In hadronic decay mode, 
The branching ratio of hadronic decay mode is as large as around 45 \%.
However, it suffers from large multi-jet background. 
In this analysis, 6 jets and at least two b-tagged jets are required.
ATLAS performed the analysis with the integrated luminosity of \mbox{1.08 fb$^{-1}$}~\cite{CERNATLAShadronic}.
Additionally $E^{miss}_{T}$ significance of \mbox{$E^{miss}_{T}$/$\sqrt{H_T}$ $<$ 3} is applied. \mbox{$\Delta R(b,\bar{b})$ $>$ 1.2} 
is also applied to remove gluon splitting.     
The event mixing technique is used modeling higher jet multiplicity using lower jet 
multiplicity multi-jet sample. The number of signal is extracted from fitting to mass $\chi^2$.   
Measured cross section is found to be
\begin{center}
$\sigma_{t\bar{t}}$=167$\pm$18(stat.)$\pm$78(syst.)$\pm$6(lumi.) pb.
\end{center}
CMS performed the analysis with the integrated luminosity of \mbox{1.1 fb$^{-1}$}~\cite{CERNCMShadronic}.
The multi-jet shape is obtained from data extrapolating from non b-tagged jet sample
(more than 6 jets) to b-tagged jets. In order to take into account the kinematic phase space difference,
the scale factor is applied to non b-tagged jet sample as a function of $p_T$ and $\eta$.
Unbinned maximum likelihood fitting is applied to top mass distribution to extract the number of signal.
Result of the fit to the reconstructed top mass is shown in Fig.~\ref{fig:hadronic}.
Measured cross section is found to be
\begin{center}
$\sigma_{t\bar{t}}$=136$\pm$20(stat.)$\pm$40(syst.)$\pm$8(lumi.) pb.
\end{center}
The uncertainties are mainly from b-tagging, JES, multi-jet background estimation.
\subsection{Combined result}
\label{sec:comb}
\begin{figure}
\begin{center}
\resizebox{0.75\columnwidth}{!}{%
\includegraphics{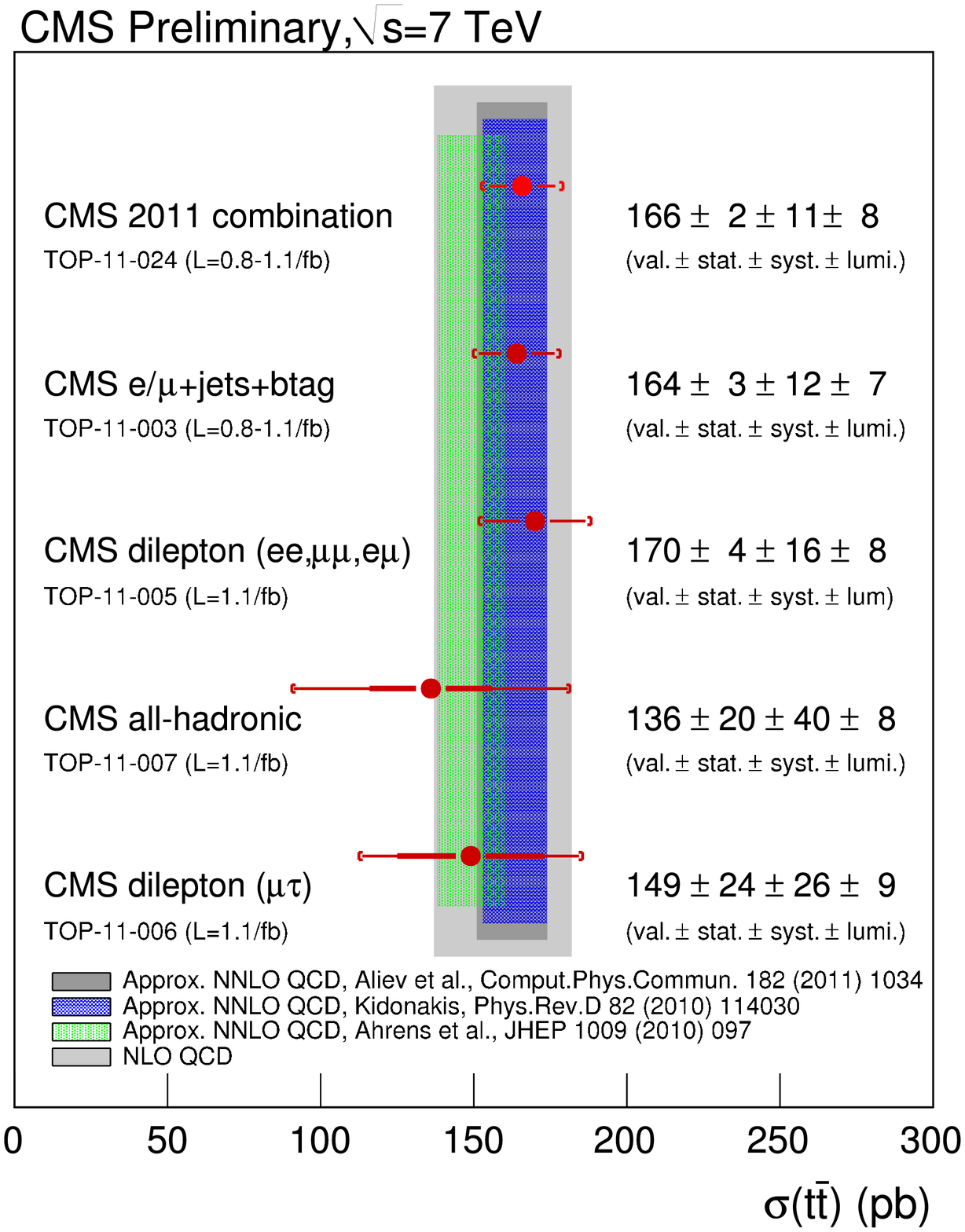} }
\caption{Combined cross section result at CMS. The data are compared to the approximate NNLO calculations
performed using the pole mass of top quark \mbox{m$^{pole}_{t} = 172.5$ GeV/c$^2$}.}
\label{fig:CMScombined}       
\end{center}
\end{figure}

ATLAS has shown the combined result from the dilepton analysis performed with \mbox{0.7 fb$^{-1}$} and lepton+jets
analysis performed with 35 pb$^{-1}$~\cite{ATLAS-CONF-2011-108}.
The combined result is found to be
\begin{center}
$\sigma_{t\bar{t}}$=176$\pm$5(stat.)$\pm$$^{13}_{10}$(syst.)$\pm$7(lumi.) pb.
\end{center}
At the time of HCP2011, 
CMS combined the dilepton (ee, e$\mu$, $\mu\mu$, $\mu\tau$), lepton+jets (e, $\mu$) and all hadronic decay analysis 
performed with around 1 fb$^{-1}$~\cite{CMS-PAS-TOP-11-024}.
The binned maximum likelihood fitter from lepton+jet analysis is used for combination
adding other decay modes as a single bin. 
Combined cross section result at CMS comparing to the approximate NNLO calculations are shown in Fig.~\ref{fig:CMScombined}.
The combined result is found to be
\begin{center}
$\sigma_{t\bar{t}}$=165.8$\pm$2.2(stat.)$\pm$10.6(syst.)$\pm$7.8(lumi.) pb.
\end{center}
The total uncertainty in combined analysis at CMS only is obtained to be 8\%,
which is the most precise measurement at the LHC. 

\section{Jet multiplicity}
\label{sec:jet}
\begin{figure}
\begin{center}
\resizebox{0.75\columnwidth}{!}{%
\includegraphics{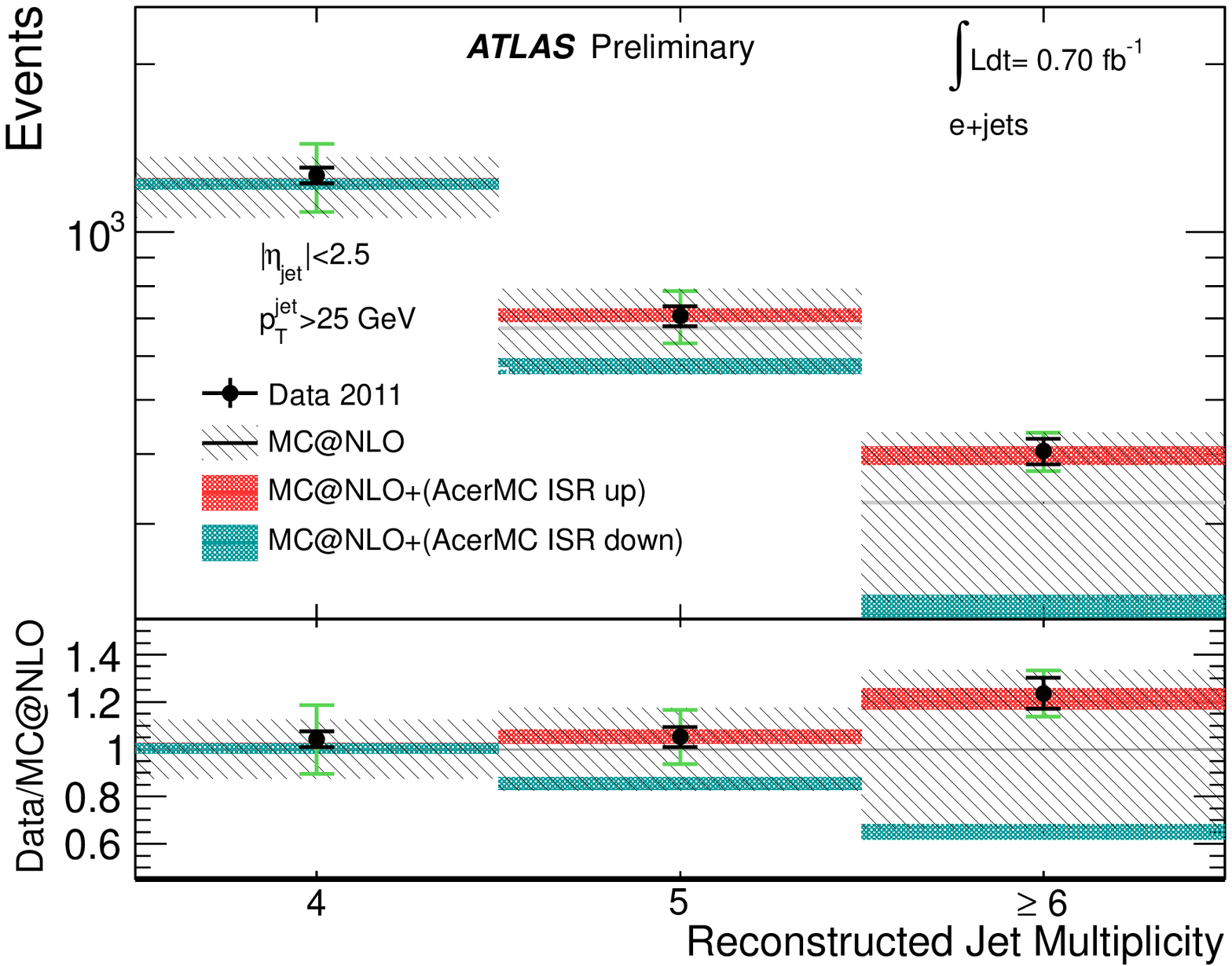}  
}
\resizebox{0.75\columnwidth}{!}{%
\includegraphics{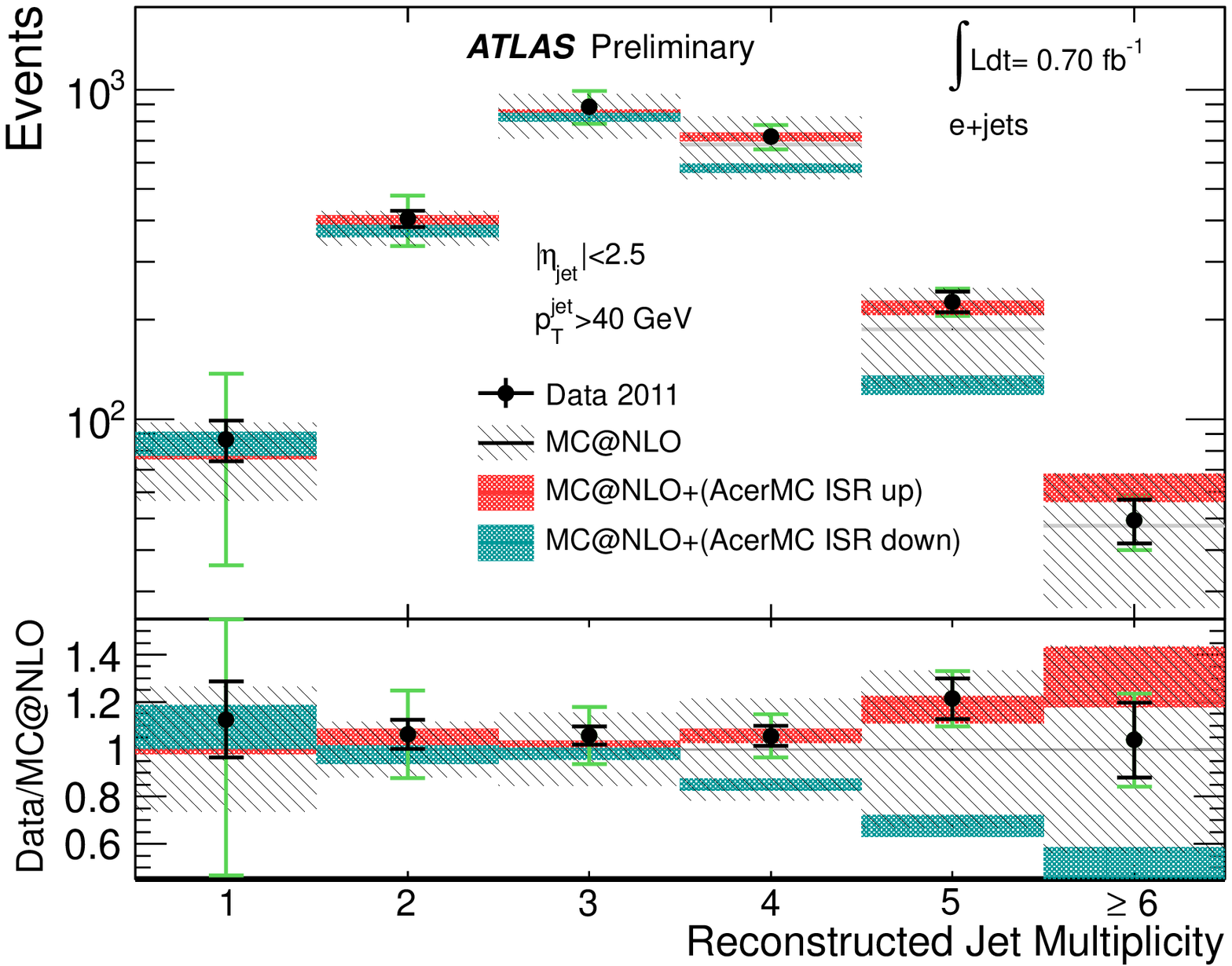} 
}
\resizebox{0.75\columnwidth}{!}{%
\includegraphics{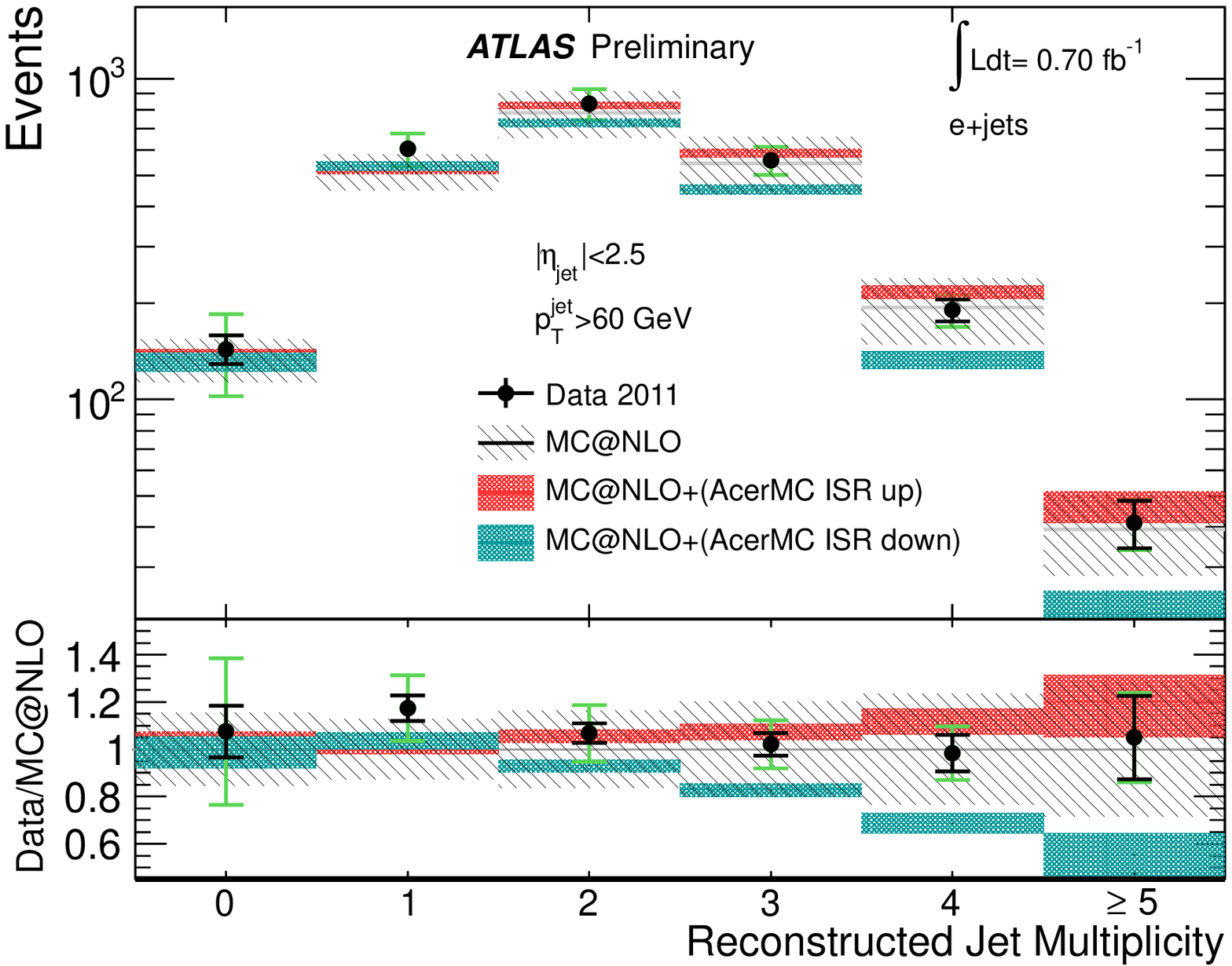} 
}
\caption{The reconstructed-jet multiplicity with different jet p$_T$ threshold (25, 40, \mbox{60 GeV}) after background subtraction at ATLAS}
\label{fig:jetATLAS}       
\end{center}
\end{figure}

Jet multiplicity distribution of $t\bar{t}$ with additional jets 
with different jet transverse momentum is very useful to constrain ISR.
ATLAS performed the analysis in lepton+jets channel with a luminosity of \mbox{0.7 fb$^{-1}$}~\cite{ATLAS-CONF-2011-142}.
Event selection follows lepton+jet analysis with requiring at least 4 jets
and one b-tagging.
The background-subtracted reconstructed jet multiplicity as a function of 
jet $p_T$ threshold (25, 40 and \mbox{60 GeV}) is 
compared with ISR variations as shown in Figs~\ref{fig:jetATLAS}.
The ISR variations were generated by varying the settings of the PYTHIA generator.
There is no deviation found from MC@NLO SM prediction.
We need more statistics to constrain ISR. 
\section{Charge asymmetry measurement}
\label{sec:charge}
\begin{figure}
\begin{center}
\resizebox{0.95\columnwidth}{!}{%
\includegraphics{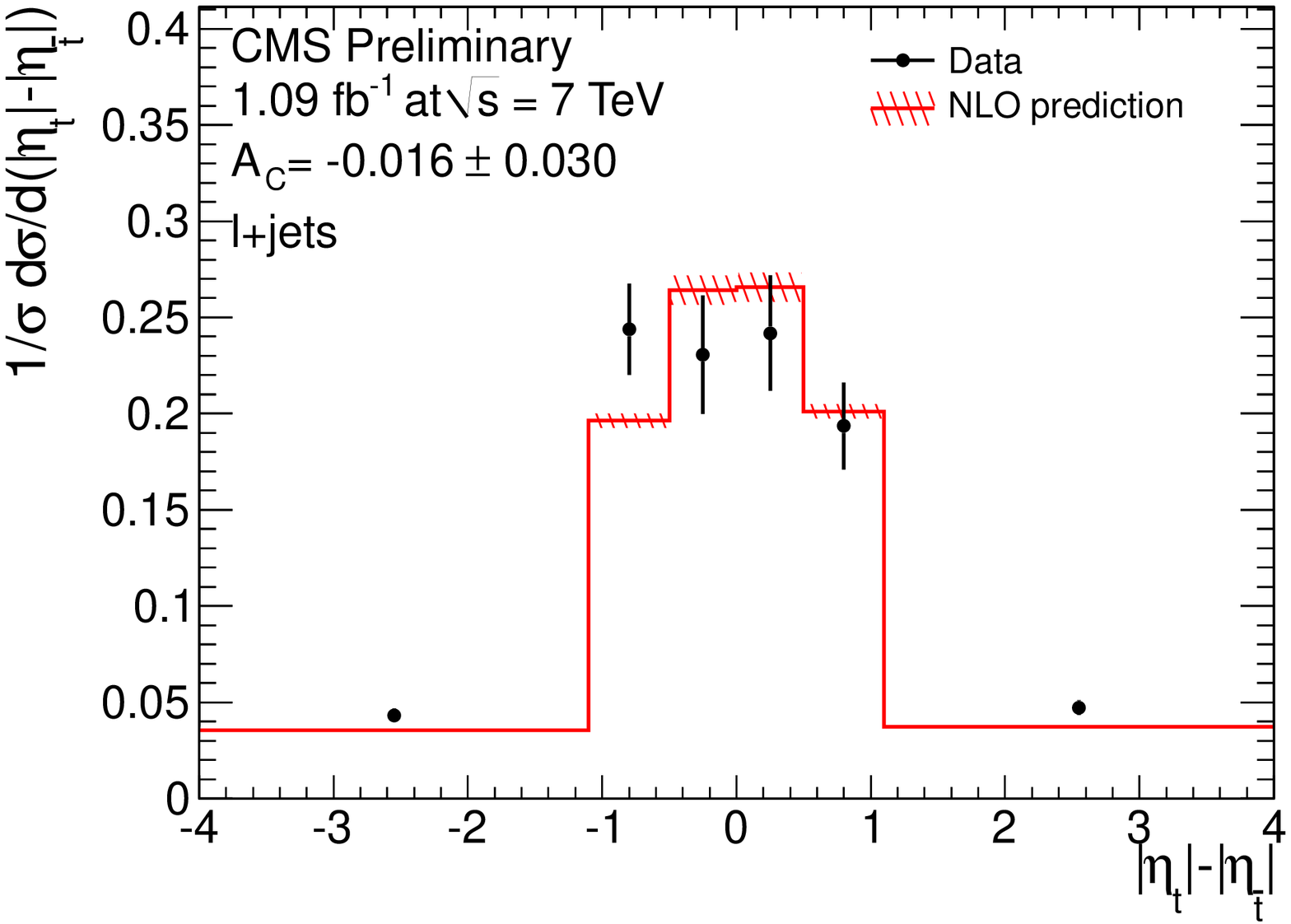}
\includegraphics{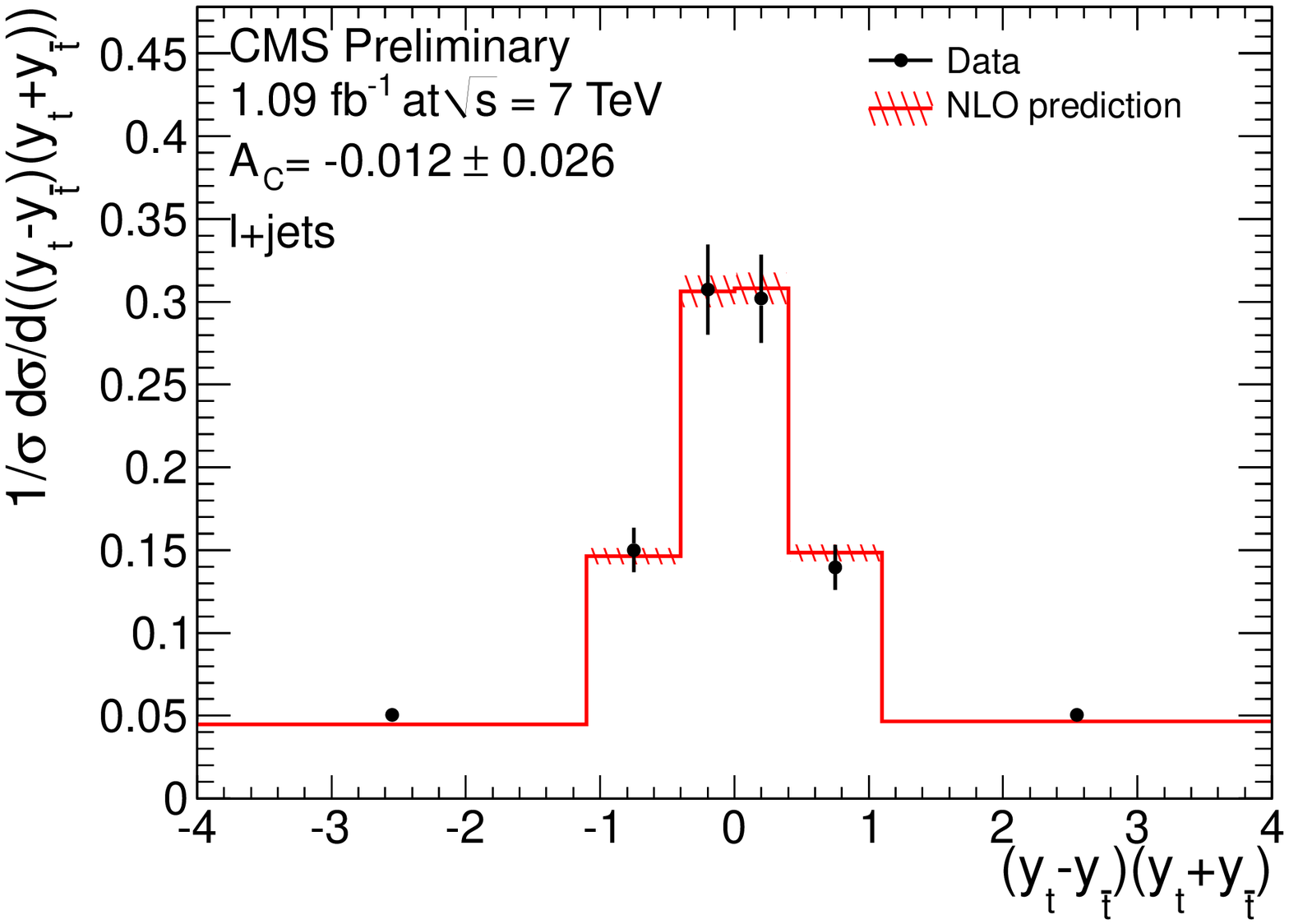}
}
\resizebox{0.95\columnwidth}{!}{%
\includegraphics{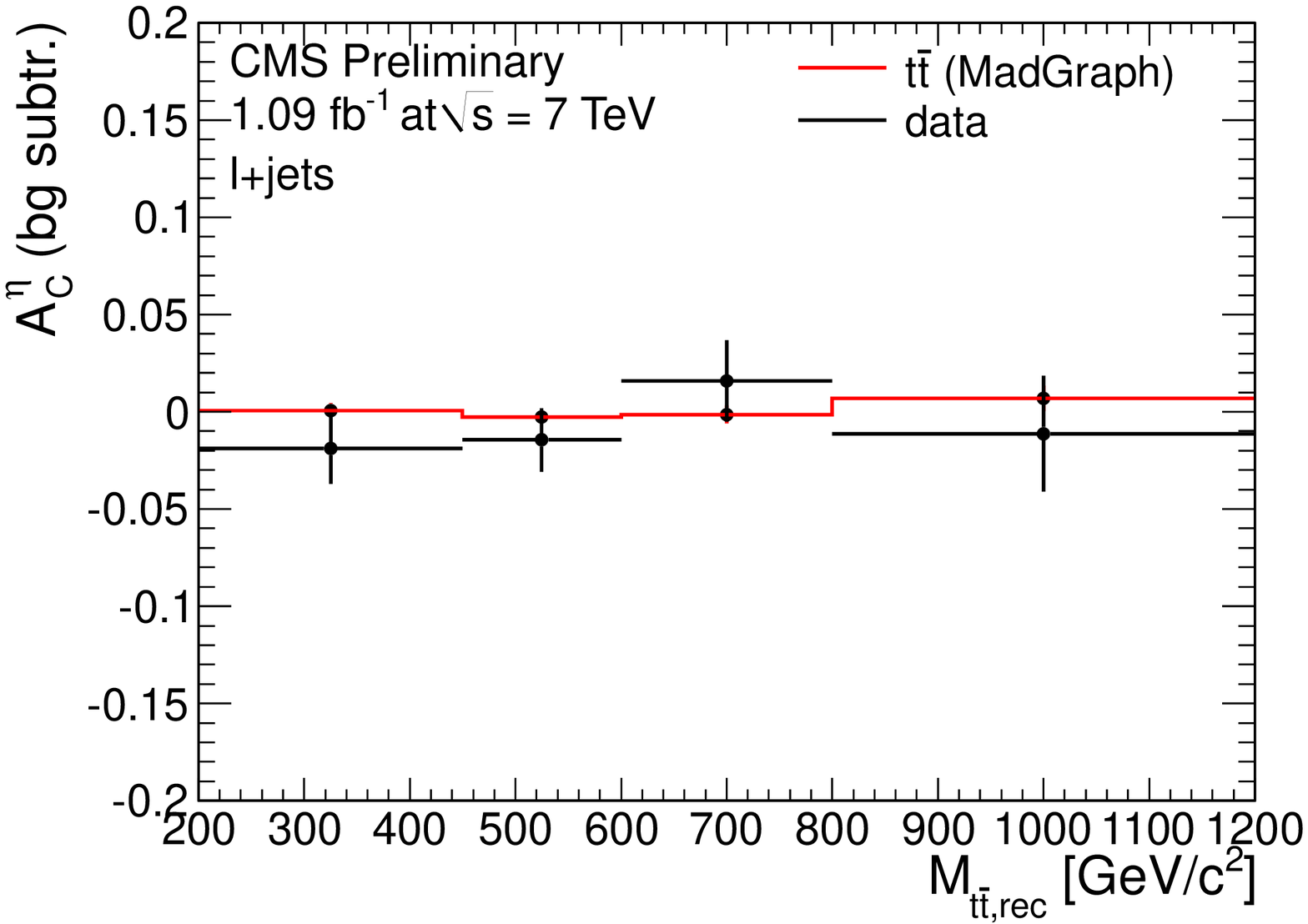} 
\includegraphics{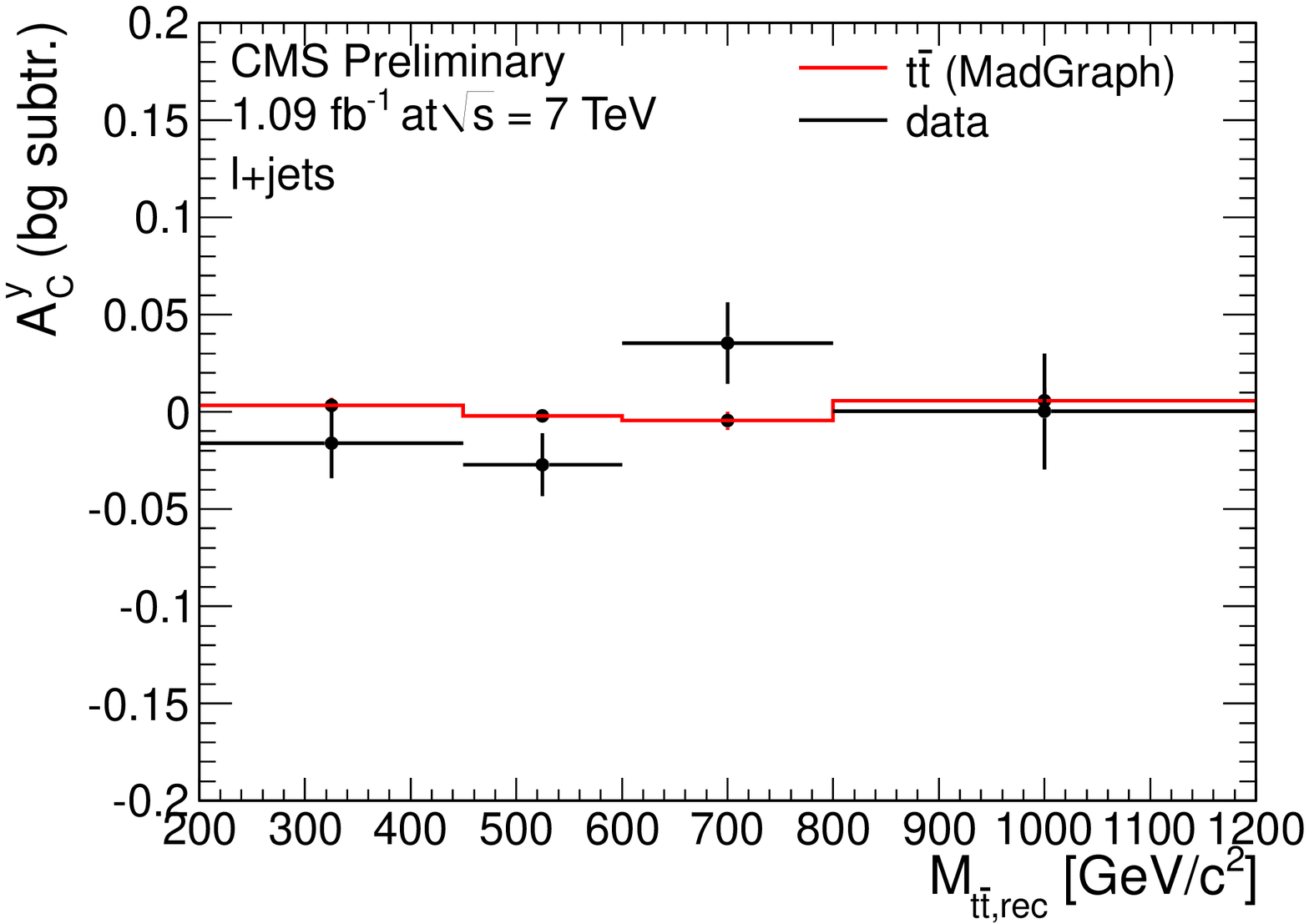} 
}
\caption{Unfolded $\Delta (|\eta|)$ spectrum (upper-left) and unfolded $\Delta (|y^2|)$ spectrum (upper-right).
The NLO SM prediction is also shown. Raw charge asymmetries for $\Delta (|\eta|)$ (lower-left) and $\Delta (|y^2|)$ (lower-right) 
as a function of reconstructed $t\bar{t}$ system mass at CMS.}
\label{fig:CMSca}       
\end{center}
\end{figure}

Tevatron has observed deviation in charge asymmetry measurement.
CDF has observed 3.4 $\sigma$ deviation with respect to SM above \mbox{450 GeV}.
The deviation could be explained by possible new exchange particles in t-channel from various theories.
The charge asymmetry variable is sensitive to this additional production mode.
Unlike Tevatron, it is hard to measure in pp collision due to the symmetric initial state of pp collisions at the LHC.
Therefore, the rapidity distributions of top and anti-top quarks are symmetrically distributed around zero. 
However, it is feasible when we consider top quark (valence quark) width is broader than anti-top quark (see quark) width.
At ATLAS, absolute rapidity is used as charged asymmetry variables 
for the analysis with an integrated luminosity of \mbox{0.7 fb$^{-1}$}~\cite{ATLAS-CONF-2011-106} while at CMS 
absolute pseudo-rapidity $\Delta (|\eta|)$ = $|\eta_t| - |\eta_{\bar{t}}| $
and rapidity with boosted effect $\Delta (y^2)$ = $( y_t - y_{\bar{t}}) \times ( y_t + y_{\bar{t}}) $ are used 
for the analysis with an integrated luminosity of \mbox{1.1 fb$^{-1}$}~\cite{CMS-PAS-TOP-11-014}.
Event selection follows lepton+jets analysis requiring 4 jets and one b-tagging.
Regularized unfolding method is applied.
Additionally the CMS performed the charge asymmetry as a function of $t\bar{t}$ reconstructed system mass 
to see the deviation above \mbox{450 GeV} (see Fig.~\ref{fig:CMSca}).
The measured charge asymmetry values at both ATLAS and CMS are within the uncertainties
in agreement with the SM theory predictions.
However, 2D unfolding method is required to confirm the deviation in high mass range.

\section{Conclusion}
\label{sec:conclusion}
We have produced precise measurement dilepton and lepton+jets decay modes both at ATLAS and at CMS. 
These measurements are already systematically limited and starting to constrain theory.
Improve pileup modeling and b-tagging is required to reduce the systematic uncertainty.
The first measurements in fully hadronic decays and decays tau are presented.
All measured results are compatible with SM prediction so far.
%

\end{document}